\documentclass[a4paper,12pt]{article}
\usepackage{authblk}
\usepackage[margin=1in]{geometry}
\usepackage{titling}
\usepackage{url}
\usepackage{amsmath, amssymb}
\usepackage{graphicx}
\usepackage{bm}
\usepackage{tikz, pgfplots}
\pgfplotsset{compat=1.18}
\usepackage{color, xcolor}
\usepackage[table]{xcolor}
\newcommand{\cg}{\cellcolor[HTML]{C0C0C0}}
\newcommand{\cb}{\color[HTML]{3166FF}}
\usepackage{float}
\usepackage{natbib}
\usepackage{rotating}
\usepackage{pdflscape}
\usepackage[normalem]{ulem}
\usepackage{listings}
\usepackage[doublespacing]{setspace}
\usepackage[style=british]{csquotes}
\usepackage{multirow, multicol}
\usepackage{booktabs}
\usepackage{tabularx}
\usepackage{siunitx}
\usepackage{subcaption} 
\usepackage{algorithm, algorithmic}
\usepackage{pgfgantt}
\usepackage{array}
\usepackage{pdfpages}
\usepackage[font=scriptsize]{caption}
\usepackage{titlesec}
\usepackage{threeparttable}

\usepackage{hyperref}

\hypersetup{
    colorlinks=true,
    linkcolor=blue, 
    citecolor=blue, 
    filecolor=magenta, 
    urlcolor=blue 
}

\title{\Large Global Stock Market Volatility Forecasting Incorporating Dynamic Graphs and All Trading Days}
\author{Zhengyang Chi\thanks{Corresponding author. Email: zhengyang.chi@sydney.edu.au}}
\author{Junbin Gao}
\author{Chao Wang\thanks{Email for all authors: \{zhengyang.chi, junbin.gao, chao.wang\}@sydney.edu.au}}

\affil{\small Discipline of Business Analytics, The University of Sydney Business School}

\begin{document}
\date{}
\maketitle

\begin{abstract}
This paper introduces a global stock market volatility forecasting model that enhances forecasting accuracy and practical utility in real-world financial decision-making by integrating dynamic graph structures and encompassing all active trading days of different stock markets. The model employs a spatial-temporal graph neural network architecture to capture the volatility spillover effect, where shocks in one market spread to others through the interconnective global economy. By calculating the volatility spillover index to depict the volatility network as graphs, the model effectively mirrors the volatility dynamics for the chosen stock market indices. In the empirical analysis covering 8 global market indices, the realized volatility forecasting performance of the proposed model surpasses the baseline models in all forecasting scenarios.
\end{abstract}


\section{Introduction}
Financial volatility forecasting is critical in financial applications, including risk management, portfolio allocation and option pricing. Although volatility cannot be directly observed because it involves the inherent variability of the returns over a time period, different types of volatility measurements have been proposed to estimate the volatility \citep{Wilmott2013}. For instance, the Realized Volatility (RV), proposed by \citet{Andersen1998}, uses the sample track of the intraday return with sufficiently high sampling frequency to assess the daily volatility.

Various models have been proposed to forecast RV. Among them, the Heterogeneous Auto-Regressive (HAR) model \citep{Corsi2009} becomes one of the most frequently used models due to its simplicity and forecast accuracy. However, it does not consider the volatility spillover effect, the co-movement and correlation among the volatility of different assets, when forecasting RV. The volatility spillover effect has been recognized as a salient attribute of financial volatility and has been frequently discussed in the literature, e.g., \citet{Kanas2000}, \citet{Forbes2002}, \citet{Poon2003}, \citet{Diebold2009}, \citet{Yang2017}, and \citet{Bollerslev2018}. For the global stock market volatility forecasting task, the volatility spillover effect provides the theoretical foundation on how volatility changes in one market can precipitate similar changes in others. Such interdependencies highlight the complexity of global financial systems and the need for more powerful volatility forecasting models. Although multiple extension HAR models, including the  Vector HAR (VHAR) \citep{Bubak2011} and the HAR-Kitchen Sink (HAR-KS) models \citep{Liang2020}, have been proposed to take the volatility spillover effect into account, they limit the volatility interactions to linear dependencies. Recently, graph neural networks (GNNs) based models, such as the Spatial Temporal Graph Spillover (STG-Spillover) model \citep{Son2023} and the GNN-HAR model \citep{Zhang2025}, are proposed to capture the nonlinear volatility interactions. They further improve the RV forecasting accuracy.

GNNs are a type of neural network designed to learn from data structured as graphs. They can proficiently capture the interactions between different nodes (entities) within the graphs through neural network layers that iteratively aggregate and transform information from neighboring nodes. This process allows GNNs to learn complex patterns in the graph structure and makes them suitable for tasks where relationships are important, e.g., analyzing the volatility spillover effect. In addition, many spatial-temporal GNN models are proposed to handle systems of time series such as traffic flows \citep{Bui2022}. Unlike other commonly used neural network architectures for sequential data, such as the Long Short-Term Memory (LSTM) and Transformer which primarily focus on temporal dependencies within individual time series, spatial-temporal GNN models simultaneously capture both the temporal dynamics within each series and the cross-sectional interdependencies across different series. In the settings of spatial-temporal GNNs, different time series data can be modeled as nodes in a graph with their relationships represented as edges. This enables the model to learn from both individual time series and their mutual influences, thus generating more accurate forecasts for the dynamics of interconnected systems.

Although great efforts have been made to utilize the volatility spillover effect to enhance RV forecasting accuracy, there are still some potential limitations in the existing work. 
\begin{itemize}
    \item When dealing with the volatility spillover effect, all the above-mentioned RV forecasting models are assumed to be trained and generate outputs on common trading days. Uncommon trading days are removed from the dataset prior to the model training process. This can significantly limit the practical utility of the RV forecasting models when cross-market investments are interesting to investors. Models only learn from common trading days and their $h$-step ahead RV forecasts are for the next $h$ common trading days rather than the next $h$ actual trading days. Models trained through this approach miss important volatility information on days when only a subset of markets is active. For instance, during the first few days of the Lunar New Year, the Hong Kong stock market remains inactive while U.S. markets are trading. Consequently, American investors may not receive accurate and reliable volatility forecasts from the STG-Spillover or GNN-HAR models mentioned above for periods that are active for their local stock market.
    
    \item The above-mentioned RV forecasting models that consider the volatility spillover effect stick to fixed volatility spillover relational graphs. This means that no matter how long a period is selected, the interdependence pattern among different stock markets is always fixed. This could be further improved, especially in terms of volatility dynamics. Many time-varying factors could be influential to the potential volatility dynamics. For example,  market-specific trading schedules can cause the volatility interconnection pattern to vary on a daily basis. Besides, other factors, such as policy evolution, global crises, local disasters and divergent development trajectories, can lead to more periodic changes in the structure of the volatility interconnection.
\end{itemize}

This research aims to address the limitations mentioned above and to enhance the practical utility and RV forecasting accuracy of global stock markets. Given the flexibility and expressiveness requirement for the desired model, this research employs the Diffusion Convolutional Recurrent Neural Network (DCRNN) model \citep{Li2018}, which is a powerful model that can learn from both relational and sequential data. The main contributions of the paper are summarized below.
\begin{itemize}
    \item This research is the first to explicitly take both common and uncommon trading days into consideration when performing global stock market volatility forecasting. The proposed model is equipped with specifically designed data masks to accommodate trading schedule differences of various global stock markets. It offers a dynamic solution that adapts to variations in different market trading patterns, thus enhancing its practical utility and forecasting reliability.
    \item The proposed model creatively combines the DCRNN model and the HAR model to generate more accurate RV forecasts. The proposed model is named DCRNN-HAR. Compared to the traditional HAR family of models, the DCRNN-HAR model can account for the nonlinear volatility spillover effect, to more flexibly capture the complex volatility dynamics. Compared to the STG-Spillover model, the proposed model leverages the dynamic volatility spillover graph and HAR framework to enhance forecasting accuracy. Compared to the GNN-HAR model which relies on a static Graph Convolutional Network (GCN), the dynamic volatility network is captured in the DCRNN-HAR.
\end{itemize}

This paper is structured as follows. Section \ref{sec:literature_review} reviews of the relevant literature. Section \ref{sec:methodology} elaborates on the methods and practical implementation of the proposed model. Detailed information about the experiments is presented in Section \ref{sec:empirical_study}, and the relevant codes are accessible at {\small\url{https://github.com/MikeZChi/DCRNN-HAR.git}}. Finally, Section \ref{sec:conclusion} concludes the paper.

\section{Literature Review}\label{sec:literature_review}
Given that this research focuses on RV forecasting, discussions on volatility forecasting methods are limited to relevant RV-focused models only. The HAR model, proposed by \citet{Corsi2009} and improved by \citet{Bollerslev2018}, is one of the most commonly used models in RV forecasting. The HAR model also has multiple extensions, which are covered in this section.

The HAR model identifies the overall pattern of volatility across three distinct time intervals. It uses the pooled panel data consisting of the past daily, weekly and monthly RV to forecast the future RV. These three inputs are expected to reflect the short-term, mid-term and long-term behaviors of the investors, respectively. The univariate HAR model for individual stock market $i$ ($i=1, 2, ..., N$) can be formulated as:
\begin{equation} \label{HAR}
    \text{RV}_{i,t} = \alpha_i + \beta_{i,d} \text{RV}_{i,t-1} + \beta_{i,w} \text{RV}_{i,t-5:t-1} + \beta_{i,m} \text{RV}_{i,t-22:t-1} + \epsilon_{i,t}, \epsilon_{i,t} \sim N(0, \sigma_{\epsilon_i}^2),
\end{equation}
where $\text{RV}_{i, t-n:t-1} = \frac{1}{n}\sum_{j = t-n} ^ {t-1} \text{RV}_{i,j}$ is the mean RV of the $i^\text{th}$ stock market from time $t-1$ to time $t-n$. Here, the $\beta_{i,d}$, $\beta_{i,w}$ and $\beta_{i,m}$ are scalars which, for each stock market, represent the past daily, weekly and monthly impact on forecasting future RVs. Besides, the intercept term $\alpha_i$ is also a scalar. The log-transformed version and the square root of the RV data are commonly used in the HAR model to deliver better forecasting performance, because the original RV data typically exhibits skewness and leptokurtosis (fat tails). Although the forecasting capability can be improved with the transformed version of RV, the univariate HAR model does not capture the relationships between the volatility of different stock markets. It forecasts the RV of each observed stock market in an isolated manner, which can be improved by considering the salient phenomenon of the volatility spillover effect.

Suppose $\mathbf{x}_t = [\text{RV}_{1,t}^\frac{1}{2}, \text{RV}_{2,t}^\frac{1}{2}, \ldots, \text{RV}_{N,t}^\frac{1}{2}] \in \mathbb{R}^N$ is used to denote the RV observations for all the $N$ market indices at time $t$. Proposed by \citet{Bubak2011}, the Vector HAR (VHAR) model captures the relationship among the RV panel data of several European foreign exchange markets. The model can be formulated in the following: 
\begin{equation} \label{VHAR}
    \mathbf{x}_t = \boldsymbol{\alpha} + \boldsymbol{\beta}_d \mathbf{x}_{t-1} + \boldsymbol{\beta}_w \mathbf{x}_{t-5:t-1} + \boldsymbol{\beta}_m \mathbf{x}_{t-22:t-1} + \boldsymbol{\epsilon}_t, \boldsymbol{\epsilon}_t \sim N(0, \boldsymbol{\Sigma}_{\epsilon}^2),
\end{equation}
where $\boldsymbol{\beta}_d$, $\boldsymbol{\beta}_w$ and $\boldsymbol{\beta}_m$ are $\mathbb{R}^{N \times N}$ matrices to capture the interplay between different markets. These square matrices of trainable parameters allow the model to learn the joint behavior of the RV data in different stock markets.

The HAR-KS model is proposed by \citet{Liang2020} to incorporate the relational information between stock markets into the univariate HAR model. The `KS' represents Kitchen Sink, which means that the model includes a variety number of features. Specifically, the HAR-KS adds the past daily RV data of other stock market indices as additional variables to the HAR model. The HAR-KS employs the square root of the RV as well. The forecasting model for each individual stock market $i$ can be formulated below:
\begin{equation} \label{HAR-KS}
\begin{split}
    \left(\text{RV}_{i,t}\right)^{1/2} &= \beta_{i,0} + \beta_{i,d} \left(\text{RV}_{i,t-1}\right)^{1/2} 
        + \beta_{i,w} \left(\text{RV}_{i,t-5:t-1}\right)^{1/2} \\
        &+ \beta_{i,m} \left(\text{RV}_{i,t-22:t-1}\right)^{1/2} 
        + \sum_{j \in \{1,\ldots,N\} \setminus \{i\}} \beta_{j,d} \left(\text{RV}_{j,t-1}\right)^{1/2} + \varepsilon_{i,t},
\end{split}
\end{equation}
Through adding extra features and coefficients, the relationship between the different stock markets can be learned through training. However, the VHAR model and the HAR-KS model both assume the volatility interdependence between stock markets is linear.

The GNN-HAR is proposed by \citet{Zhang2025} to capture the nonlinear volatility spillover effect through a multilayer GCN framework. Instead of learning the relationship network through training, the graph structure used in the GNN-HAR model is captured by the Graphical LASSO (GLASSO) method \citep{Friedman2007} before the actual training procedure to enhance the efficiency of the model. The precision matrix derived through the GLASSO algorithm is transformed and used as the adjacency matrix for the volatility relationship graph. The GNN-HAR model directly adds its graph design to the HAR model with slight modifications on the time interval of the mid-term and long-term RV data to avoid overlapping:
\begin{equation}\label{GNN-HAR}
    \begin{split}
        \mathbf{H}^{(0)} =&\; [\mathbf{x}_{t-1}, \mathbf{x}_{t-5:t-2}, \mathbf{x}_{t-22:t-6}],\\
        \mathbf{H}^{(1)} =&\;  \text{ReLU}(\mathbf{D}^{-\frac{1}{2}} \mathbf{A} \mathbf{D}^{-\frac{1}{2}}\mathbf{H}^{(0)}\mathbf{W}^{(0)}),\\
        & \ldots,\\
        \mathbf{H}^{(K)} =&\; \text{ReLU}(\mathbf{D}^{-\frac{1}{2}} \mathbf{A} \mathbf{D}^{-\frac{1}{2}}\mathbf{H}^{(K-1)}\mathbf{W}^{(K-1)}),\\
        \mathbf{x}_t =&\; \boldsymbol{\alpha} + \beta_d \mathbf{x}_{t-1} + \beta_w \mathbf{x}_{t-5:t-2} + \beta_m \mathbf{x}_{t-22:t-6} \\
        &\;   + \boldsymbol{\gamma} \mathbf{H}^{(K)} +\boldsymbol{\epsilon}_t,
    \end{split}
\end{equation}
where $\mathbf{A} \in \mathbb{R}^{N \times N}$ is the estimated adjacency matrix constructed based on the precision matrix from the GLASSO algorithm and $\mathbf{D} \in \mathbb{R}^{N \times N}$ is the diagonal matrix $\mathbf{D} = \text{diag}(d_1, \ldots, d_N)$, in which $d_i = \sum_{j = 1}^{N} A_{ij}$. $\mathbf{H}^{(0)} \in \mathbb{R}^{N \times 3}$ is the pooled panel RV data. Besides, $\beta_d$, $\beta_w$ and $\beta_m$ are the impact of the lagged panel RV observations similar to the HAR model, whereas $\boldsymbol{\gamma} = [\gamma_d, \gamma_w, \gamma_m]$ measures the influence from the past RV values from the neighborhood entities. $\text{ReLU}(\cdot)$ is the nonlinear activation function to capture the nonlinear volatility spillover effect. $K$ is the number of GCN lays, which measures the influential range of the volatility spillover effect. $\{\mathbf{W}^{(k)}\}_{k=0}^{K-1}$ are learnable parameters in GCN.

A spatial-temporal GNN model, DCRNN, proposed by \citet{Li2018} is leveraged to enhance the RV forecasting accuracy \citep{Son2023}. This model is originally designed to address the challenges in traffic forecasting by integrating cross-sectional and temporal dependencies. The DCRNN model uses diffusion convolution to model cross-sectional dependencies as a diffusion process on a directed graph and employs Gated Recurrent Units (GRUs) as a variant of typical recurrent neural networks (RNNs) to capture temporal dynamics of traffic flow. The model also applies the encoder-decoder architecture for improved long-term forecasting. This model is applied to RV forecasting on 8 different stock markets and achieves satisfactory results compared to the linear HAR models \citep{Son2023}. Similar to the GNN-HAR model, the volatility relational graph is constructed before the training stage. The volatility spillover effect is captured by the volatility spillover index under the Diebold \& Yilmaz (DY) framework proposed by \citet{Diebold2012}. Thus, in the RV forecasting scenario, the model is referred to as the STG-Spillover model \citep{Son2023}. The details of the DY framework and the DCRNN model are discussed in later sections, because these components are also applied in the proposed model of this research.

\section{Methodology and Implementation}\label{sec:methodology}
This section introduces the detailed design of the proposed model, including the basic graph learning theory, the strategy to handle uncommon trading days, the method to learn the volatility graph adjacency matrix from data, and the formulation of the proposed DCRNN-HAR model. 

\subsection{Graph Learning for Relational Data}
Suppose the graph $\mathcal{G} = (\mathcal{V}, \mathcal{E})$ consists of $N$ nodes, where $\mathcal{V}$ and $\mathcal{E}$ denote the set of nodes and the set of edges in the graph, respectively. Furthermore, $v_i \in \mathcal{V}$ represents the $i^\text{th}$ node within the node set and $e_{ij} = (v_i,v_j) \in \mathcal{E}$ denotes the edge between node $v_i$ and node $v_j$ within the edge set. Here, matrix $\mathbf{Z} \in \mathbb{R}^{N \times D}$ can be used to denote the collection of the features of all nodes. Especially, the $i^\text{th}$ row of matrix $\mathbf{Z}$ is the feature vector of node $v_i$. In addition, the adjacency matrix $\mathbf{A}$ and degree matrix $\mathbf{D}$ can be used to measure the relationships or interactions between nodes and store the structural information of the corresponding graph. For example, consider $\mathcal{G}$ as a simple undirected and unweighted graph. Both the adjacency matrix $\mathbf{A} \in \mathbb{R}^{N \times N}$ and the degree matrix $\mathbf{D} \in \mathbb{R}^{N \times N}$ are square matrices. They are formulated in the following ways:
\begin{equation} \label{adj mx}
    \mathbf{A}_{ij} =
    \begin{cases}
    1, &\mbox{$(v_i, v_j) \in \mathcal{E}$};\\
    0, &\mbox{$(v_i, v_j) \notin \mathcal{E}$},
    \end{cases}
\end{equation}
and
\begin{equation}\label{deg mx}
    \mathbf{D} = \text{diag}(\sum_{j = 1}^{N} \mathbf{A}_{1j}, \ldots, \sum_{j = 1}^{N} \mathbf{A}_{Nj})
\end{equation}

For the adjacency matrix $\mathbf{A}$, if node $v_i$ and node $v_j$ are connected, the corresponding element $\mathbf{A}_{ij}$ is set to $1$. Otherwise, $\mathbf{A}_{ij}$ is $0$ meaning node $v_i$ and $v_j$ are not connected. The degree matrix $\mathbf{D}$ is a diagonal matrix whose diagonal elements are the row sum of the corresponding rows of the adjacency matrix $\mathbf{A}$.

Based on the basic knowledge about the graph, the graph propagation mechanism can be described as follows. For each target node $v_i$, the information (i.e., node features) in the neighborhood of $v_i$, which is denoted as $N(v_i)$, is transformed, aggregated and combined with the transformed information of $v_i$ to update the node features of $v_i$. Suppose the information transformation function is $f(\cdot)$ and the aggregation function is $\text{AGG}(\cdot)$. The iterative node feature aggregation and update process can be depicted in the formula below:
\begin{equation}\label{spatial graph general}
    \mathbf{H}_i^{(k)} = \text{AGG}^{(k)}(\{f^{(k)}(\mathbf{H}_j^{(k-1)}), v_j \in N(v_i)\},f^{(k)}(\mathbf{H}_i^{(k-1)})),
\end{equation}
where $\mathbf{H}_i^{(0)}$ and $\mathbf{H}_j^{(0)}$ are the original node features of node $v_i$ and $v_j$, respectively. $K$ is the number of layers and represents the receptive field of message passing. The information transformation $f(\cdot)$ is conducted on node features individually, while aggregation $\text{AGG}(\cdot)$ is performed by merging information according to node neighborhood, which is defined by the adjacency matrix. Particularly, the aggregation operation must be invariant to node permutation. A simple example is formulated below.
\begin{equation}\label{spatial graph mx}
    \mathbf{H}^{(k+1)} = \sigma(\mathbf{A}\mathbf{H}^{(k)}\mathbf{W}^{(k)}),
\end{equation}
where $\mathbf{H}^{(0)}$ is the original node features $\mathbf{Z}$, $\mathbf{W}^{(l)}$ and $\sigma(\cdot)$ transform the information in linear and nonlinear manners, respectively, and the aggregation function is simply the summation among neighbor nodes.

\subsection{RV Data Handling for Uncommon Trading Days}
In order to leverage the data on both common trading days and uncommon trading days, special data masks should be designed to identify inactive and active markets at each trading day. These masks should be applied to data in both the look-back window and the forecast window and to the adjacency matrix of the volatility interconnection graph. Given the task of the DCRNN-HAR model is to forecast future RV based on historical RV, each node (stock market index) only has a scalar feature, the historical RV value, at each trading day. However, for the uncommon trading days, there are missing RV values of the inactive markets. This gives rise to the data mask at the data processing stage.

Under the HAR framework, the look-back window $l = 22$ and the forecast window $h$ can take values from $\{1, 5, 22\}$ to represent the short-term, mid-term or long-term forecasting task. Here, suppose each data point consists of a pair of the look-back RV input $\mathbf{X} \in \mathbb{R}^{l \times N}$ and the target future RV $\mathbf{Y} \in \mathbb{R}^{h \times N}$. Here, unless otherwise stated, the row index for dates is denoted by $t$ and the column index for stock market indices is denoted by $n$. In each column of $\mathbf{X}$ and $\mathbf{Y}$, there are some empty entries due to the trading date scheduling difference in $N$ distinct stock markets. Hence, mask matrices $\mathbf{E}^X \in \mathbb{R}^{l \times N}$ and $\mathbf{E}^Y \in \mathbb{R}^{h \times N}$ are constructed to handle missing values within $\mathbf{X}$ and $\mathbf{Y}$, respectively. The masks are formulated as below,
\begin{equation} \label{EX}
\mathbf{E}^X_{t n} =
    \begin{cases}
    0, &\mbox{$\mathbf{X}_{t n} \text{ is missing}$};\\
    1, &\mbox{$\mathbf{X}_{t n} \text{ exists}$},
    \end{cases}
\end{equation}
and
\begin{equation} \label{EY}
    \mathbf{E}^Y_{t n} =
    \begin{cases}
    0, &\mbox{$\mathbf{Y}_{t n} \text{ is missing}$};\\
    1, &\mbox{$\mathbf{Y}_{t n} \text{ exists}$}.
    \end{cases}
\end{equation}
The masks $\mathbf{E}^X$ and $\mathbf{E}^Y$ are equivalent to filling the missing values with $0$ at the data processing stage. They mark the distribution of active and inactive markets for each union trading day. Hence, each data point becomes $(\tilde{\mathbf{X}}, \tilde{\mathbf{Y}}) = (\mathbf{E}^X \odot \mathbf{X}, \mathbf{E}^Y \odot \mathbf{Y})$, where $\odot$ represents the element-wise multiplication between the two matrices. The mask $\mathbf{E}^X$ is important in loading inputs into the DCRNN-HAR model so that the input data can be processed by the model. $\mathbf{E}^Y$ is also important in masking RV forecasts for inactive markets so that the model can proactively ignore those meaningless outputs and behave like a stock market participant. In practice, stock exchanges in different markets release public holiday calendars annually which inform all scheduled market closures. Although unscheduled halts are unpredictable, scheduled closures and policy changes are typically communicated well in advance. Hence, it is feasible to assume that stock market participants know future trading schedules for all stock markets. In addition, masking out irrelevant RV forecasts for inactive stock markets also helps with model training. In the loss function calculation, the data mask $\mathbf{E}^Y$ can ensure that forecasts on inactive markets do not affect the values of the trainable parameters through backward propagation. Hence, the loss function with respect to each data point is calculated as follows,
\begin{equation} \label{L}
    L = \frac{\sum_{t,n} \mathbf{E}^Y_{t n} \ell(\tilde{\mathbf{Y}}_{t n}, \widehat{\mathbf{Y}}_{t n})}{\sum_{t,n} \mathbf{E}^Y_{t n}},
\end{equation}
where $\widehat{\mathbf{Y}}$ is the forecast RV values produced by the DCRNN-HAR model and $\ell(\cdot)$ is the chosen loss measurement function. For example, if $\ell(\tilde{\mathbf{Y}}_{t n}, \widehat{\mathbf{Y}}_{t n}) = (\tilde{\mathbf{Y}}_{t n} - \widehat{\mathbf{Y}}_{t n})^2$, $L$ is the Mean Squared Error (MSE). If $\ell(\tilde{\mathbf{Y}}_{t n}, \widehat{\mathbf{Y}}_{t n}) = |\tilde{\mathbf{Y}}_{t n} - \widehat{\mathbf{Y}}_{t n}|$, $L$ is the Mean Absolute Error (MAE). No matter which loss function is selected, the performance is measured only when relevant markets are active.

In addition to the two data mask matrices $\mathbf{E}^X$ and $\mathbf{E}^Y$, another graph mask is designed and applied to the volatility interconnection adjacency matrix to cut the connection from inactive markets to active markets. In this way, inactive markets do not influence active markets. However, at the same time, inactive markets can still receive volatility information from active markets so that their states can still be updated in time. Here, for each data point $(\tilde{\mathbf{X}}, \tilde{\mathbf{Y}})$, one corresponding adjacency mask is constructed in the following way. Suppose the data mask matrices are combined sequentially (vertically) such that $\mathbf{E}^{\text{data}} = [\mathbf{E}^x \Vert \mathbf{E}^y] \in \mathbb{R}^{(l+h) \times N}$, where $[\cdot \Vert \cdot]$ concatenate two matrices vertically. The adjacency mask is designed for each row of $\mathbf{E}^{\text{data}}$ and is denoted as $\mathbf{E}^{A_t} \in \mathbb{R}^{N \times N}$. Its elements in the $n^\text{th}$ row ($\mathbf{E}^{A_t}_{n}$) are set to $0$ if $\mathbf{E}^{\text{data}}_{t n} = 0$. Otherwise, elements in $\mathbf{E}^{A_t}$ are set to $1$. Thus, the formulation of $\mathbf{E}^{A_t}$ is shown below:
\begin{equation} \label{EA}
    \mathbf{E}^{A_t}_{n} =
    \begin{cases}
    \mathbf{0}_N, &\mbox{$\text{ if } \mathbf{E}^{\text{data}}_{t n} = 0$};\\
    \mathbf{1}_N, &\mbox{$\text{ if } \mathbf{E}^{\text{data}}_{t n} \neq 0$},
    \end{cases}
\end{equation}
As a result, a unique masked adjacency is constructed for each $t$: $\tilde{\mathbf{A}}_t = \mathbf{E}^{A_t} \odot \mathbf{A}$, where $\mathbf{A}$ is the adjacency matrix constructed based on each look-back window data $\mathbf{X}$ (instead of the whole in-sample data) to better capture the periodic dynamic of the volatility interconnections. More frequent changes in the volatility interconnection dynamic due to the difference in trading schedules are captured by $\mathbf{E}^{A_t}$. Although the data masks and the graph masks are straightforward, they are important to ensure that, during the training and forecasting processes, the inactive markets do not affect those active markets but can still receive relevant volatility information from active markets.

The data and graph masking process is intuitively demonstrated in Figure \ref{data_prep_demo} below. The numbers in the input $\mathbf{X}$, target $\mathbf{Y}$ and adjacency matrix $\mathbf{A}$ are randomly generated. `?' denotes inactive markets on the corresponding trading days. $\tilde{\mathbf{X}}$ and $\tilde{\mathbf{Y}}$ represent the masked input and target data matrices. The example adjacency mask $\mathbf{E}^{A_1}$ is built based on day 1 data of $\mathbf{E}^{X}$. $\tilde{\mathbf{A}}_1$ is the masked adjacency matrix for day 1 data.
\begin{figure}[H]
\centering
\includegraphics[width=\textwidth]{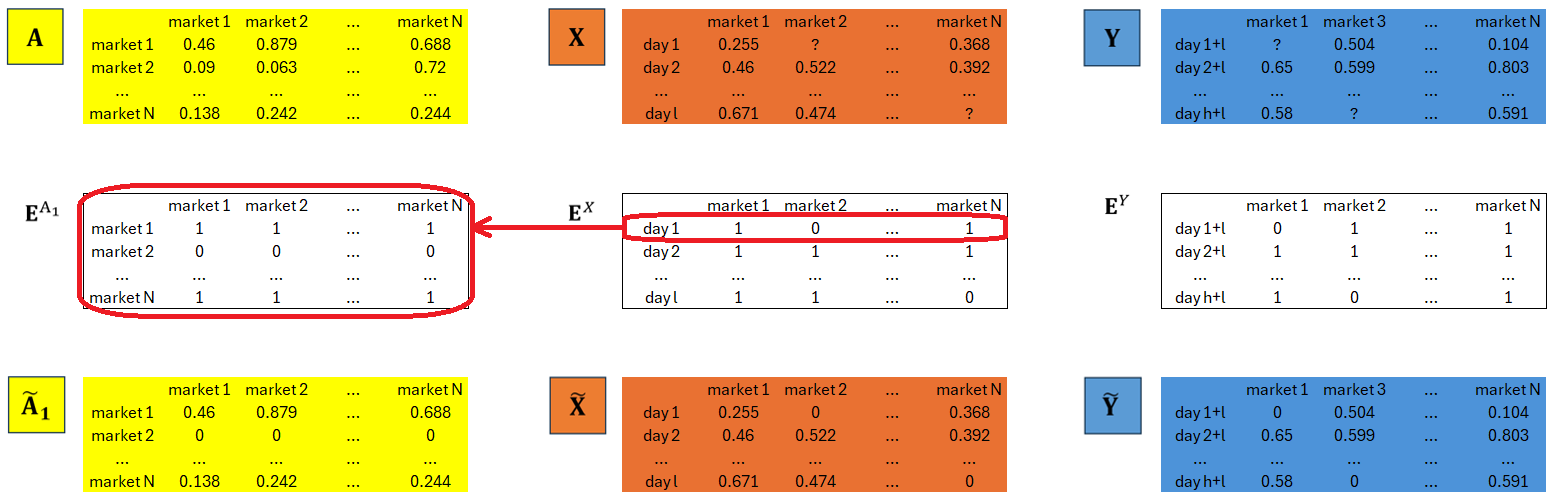}
\caption{A demonstration of how the data and graph masks are applied in the data processing stage.}
\label{data_prep_demo}
\end{figure}

\subsection{Volatility Interconnection Graph Construction}\label{sec:volatility_interconnection}
Although learning the relationships among stock markets during training may contribute to a purely data-driven model such as VHAR (Equation \eqref{VHAR}) and HAR-KS (Equation \eqref{HAR-KS}), the estimated volatility network may deviate from reality given the linear expressiveness and limited flexibility of the model. However, for powerful models such as neural networks, formulating the interactions and interrelationships among variables during training can be too costly. Hence, embedding a prespecified graph structure into the model not only brings exogenetic power to the model, but also accelerates the training process. This is also the reason that GNNs are preferred in this research. Specifically, for the DCRNN-HAR model, the volatility interconnection graph is constructed under the DY framework \citep{Diebold2012}. 

Suppose a multivariate time series $\{\mathbf{u}_t \in \mathbb{R}^N \}_{t = 1}^T$ has $N$ variables and it is covariance stationary. For this time series, a $l$-lag vector autoregression (VAR($l$)) can be formulated as:
\begin{equation}\label{VAR-p}
    \mathbf{u}_t = \sum_{i=1}^l \lambda_i \mathbf{u}_{t-i} + \boldsymbol{\epsilon}_t, \boldsymbol{\epsilon}_t \sim (0, \boldsymbol{\Sigma}_{\boldsymbol{\epsilon}}),
\end{equation}
where each $\lambda_i$ is a scalar. The equation above can be reformulated in the moving average format:
\begin{equation}\label{VAR-p-MA}
    \mathbf{u}_t = \sum_{i=1}^{\infty} \mathbf{B}_i \boldsymbol{\epsilon}_{t-i},
\end{equation}
where $\mathbf{B}_i = \sum_{j=1}^{l} \lambda_j \mathbf{B}_{i-j}$. Here, $\mathbf{B}_0 = \mathbf{I}_{N \times N}$ and $\mathbf{B}_i = 0$ if $i < 0$. Through the variance decomposition of the $h$-step-ahead forecast error, the resulting volatility spillover index matrix $\boldsymbol{\theta}^g(h)$ is derived and shown below.
\begin{equation}\label{var-decomp}
    \boldsymbol{\theta}_{ij}^g(h) = \frac{\sigma_{jj}^{-1}\sum_{t=0}^{h-1}(\mathbf{e}_i' \mathbf{B}_t \boldsymbol{\Sigma}_{\boldsymbol{\epsilon}} \mathbf{e}_j)^2}{\sum_{t=0}^{h-1}(\mathbf{e}_i' \mathbf{B}_t \boldsymbol{\Sigma}_{\boldsymbol{\epsilon}} \mathbf{B}_t' \mathbf{e}_i)}.
\end{equation}
Here, $\mathbf{e}_i$ is a selection vector with its $i^\text{th}$ elements equal to $1$ and $0$ elsewhere. $\sigma_{jj}$ is the standard deviation of $\epsilon_{tj}$, which is the error term of the $j^\text{th}$ autoregression equation in Equation \eqref{VAR-p}. The matrix $\boldsymbol{\theta}^g(h)$ can be further standardized so that the row sum of the standardized matrix $\tilde{\boldsymbol{\theta}}^g(h)$ is $1$, 
\begin{equation}\label{std-var-decomp}
   \tilde{\boldsymbol{\theta}}_{ij}^g(h) = \frac{\boldsymbol{\theta}_{ij}^g(h)}{\sum_{j=1}^N \boldsymbol{\theta}_{ij}^g(h)}.
\end{equation}
More details can be found in the original paper of the DY framework \citep{Diebold2012}.

In this research, $\tilde{\boldsymbol{\theta}}^g(h)$ is calculated based on inputs in the look-back window for each data point. Each element $\tilde{\boldsymbol{\theta}}_{ij}^g(h)$ indicates the volatility spillovers received by variable $i$ from variable $j$. To better fit the volatility spillover measurement into graphs, in this research, the transpose of the standardized volatility spillover index matrix $\tilde{\boldsymbol{\theta}}^g(h)$ is considered as the adjacency matrix of the volatility interrelationship graph (i.e., $\mathbf{A} = [\tilde{\boldsymbol{\theta}}^g(h)]^\top$).

\subsection{Proposed Model Formulation}
Targeting traffic flow forecast, the DCRNN model applies diffusion convolution to process graphical information, which is transformed to formulate the gate units in its recurrent design \citep{Li2018}. For a given graph $\mathcal{G} = (\mathcal{V}, \mathcal{E})$ with its adjacency matrix $\mathbf{A}$ (Equation \eqref{adj mx}) and degree matrix $\mathbf{D}$ (Equation \eqref{deg mx}), the diffusion process, which can capture the dynamics of the nodes in $\mathcal{V}$, is depicted by a random walk process on $\mathcal{G}$ with restart probability $p \in [0,1]$. For the random walk process, the state transition matrix can be calculated as $\mathbf{D}^{-1} \mathbf{A}$. It is the normalized adjacency matrix whose row sum equals $1$ and each element $(\mathbf{D}^{-1} \mathbf{A})_{ij}$ denotes the probability from node (state) $v_i$ to node (state) $v_j$. The random walk stochastic process can achieve stationarity after many transition iterations. The stationary distribution is denoted as $\mathbf{P} \in \mathbf{R}^{N \times N}$ whose elements $\mathbf{P}_{ij}$ represents the probability of the diffusion process from node $v_i$ to node $v_j$ after numerous transition iterations. The close form of the stationary distribution $\mathbf{P}$ is formulated below:
\begin{equation} \label{graph diffusion stationarity}
    \mathbf{P} = \sum^{\infty}_{k = 0} p (1-p)^k (\mathbf{D}^{-1} \mathbf{A})^k.
\end{equation}
As a result, the diffusion convolution operation for node features (i.e., graph signals) $\mathbf{X} \in \mathbb{R}^N$ and the corresponding filter $f_{\zeta}$ is formulated as:
\begin{equation} \label{graph diffusion conv}
    \mathbf{X} \ast _{\mathcal{G}(\mathbf{A},\mathbf{D})} f_{\boldsymbol{\zeta}} = \sum^{K-1}_{k = 0} \zeta_k (\mathbf{D}^{-1} \mathbf{A})^k \mathbf{X},
\end{equation}
where $\{\zeta_k\}_{k=0}^{K-1}$ represents the learnable parameters and $K$ is the parameter to control the receptive field of the graph information propagation through the diffusion convolution operation. The original DCRNN model also uses the reverse diffusion process. However, given the adjacency matrix is constructed under the DY framework in this paper, the pairwise volatility spillover effect is already bidirectional, thus eliminating the need to include the reverse diffusion process.

Based on the diffusion convolution operation, the temporal dynamics of the nodes are modeled through GRUs \citep{Chung2014}. In the DCRNN model, the Diffusion Convolutional Gated Recurrent Unit (DCGRU) cell is formulated in Equation \eqref{dcgru}, where $\mathbf{r}_t$ is the reset gate, $\mathbf{u}_t$ is the update gate and $\mathbf{C}_t$ is the candidate activation vector. They are used to update the hidden state $\mathbf{H}_t$.
\begin{equation} \label{dcgru}
\begin{split}
    \mathbf{r}_t &= \sigma_{\text{GRU}} (\boldsymbol{\zeta}_r \ast _{\mathcal{G}(\mathbf{A},\mathbf{D})} [\mathbf{X}_t, \mathbf{H}_{t-1}] \mathbf{W}_r + \mathbf{b}_r), \\
    \mathbf{u}_t &= \sigma_{\text{GRU}} (\boldsymbol{\zeta}_u \ast _{\mathcal{G}(\mathbf{A},\mathbf{D})} [\mathbf{X}_t, \mathbf{H}_{t-1}] \mathbf{W}_u + \mathbf{b}_u), \\
    \mathbf{C}_t &= \text{tanh} (\boldsymbol{\zeta}_C \ast _{\mathcal{G}(\mathbf{A},\mathbf{D})} [\mathbf{X}_t, (\mathbf{r}_t \odot \mathbf{H}_{t-1})] \mathbf{W}_C + \mathbf{b}_C), \\
    \mathbf{H}_t &= \mathbf{u}_t \odot \mathbf{H}_{t-1} + (1-\mathbf{u}_t) \odot \mathbf{C}_t,
\end{split}
\end{equation}
where $\ast _\mathcal{G}$ is the graph diffusion convolution defined in Equation \eqref{graph diffusion conv} and $\boldsymbol{\zeta}_r$, $\boldsymbol{\zeta}_u$ and $\boldsymbol{\zeta}_C$ are the learnable parameters of the relevant filters. $\mathbf{W}_r$, $\mathbf{W}_u$, $\mathbf{W}_C$, $\mathbf{b}_r$, $\mathbf{b}_u$ and $\mathbf{b}_C$ are learnable parameters as well. $\sigma_{\text{GRU}}$ is the nonlinear activation function for the reset and update gates. For simplicity, the DCGRU cell operation is denoted as:
\begin{equation} \label{dcgru-simple}
    \mathbf{H}_t = U_{\{\boldsymbol{\zeta}_r,\boldsymbol{\zeta}_u,\boldsymbol{\zeta}_C,\mathbf{W}_r,\mathbf{W}_u,\mathbf{W}_C,\mathbf{b}_r,\mathbf{b}_u,\mathbf{b}_C\}}(\mathbf{X}_t, \mathbf{H}_{t-1}, \mathbf{A}, \mathbf{D}).
\end{equation}
Different from the traditional HAR class of models, the DCRNN model can directly generate a sequence of multi-step ahead forecasts. The multi-step ahead forecast function is realized through the Sequence to Sequence (Seq2Seq) architecture proposed by \citet{Sutskever2014}. This architecture employs encoders and decoders, which both sequentially perform the DCGRU cell operation as Equation \eqref{dcgru-simple} to process inputs and generate forecasts.

In this paper, the proposed DCRNN-HAR model creatively combines the traditional HAR framework and the DCRNN model. It designs different masks to incorporate dynamic graphs and all trading days so that the accuracy and practical utility for global stock market RV forecasting can be enhanced. In the context of the dynamic RV forecast, the encoders encode the RV information in the look-back input window $\mathbf{X}$, whereas the decoders generate the forecast $\widehat{\mathbf{Y}}$. The adjacency matrix $\mathbf{A}$ of the volatility spillover graph $\mathcal{G}$ is generated based on $\mathbf{X}$ under the DY framework as discussed in Section \ref{sec:volatility_interconnection}. The graph mask is applied for both encoders and decoders to capture the changing structure of the volatility spillover graph: $\tilde{\mathbf{A}}_t = \mathbf{E}^{A_t} \odot \mathbf{A}$, where $\mathbf{E}^{A_t}$ is formulated as Equation \eqref{EA} and $t \in \{1,2, \ldots, l+h\}$. $\tilde{\mathbf{D}}_t$ is the corresponding degree matrix of the masked adjacency matrix $\tilde{\mathbf{A}}_t$. The diffusion convolution operation for both encoders and decoders are described in Equation \eqref{graph diffusion conv} with respect to the corresponding dynamic volatility spillover graph $\mathcal{G}(\tilde{\mathbf{A}}_t,\tilde{\mathbf{D}}_t)$. For encoders, $t \in \{1,2, \ldots, l\}$. $\tilde{\mathbf{X}} = \mathbf{E}^X \odot \mathbf{X}$ is the masked inputs (Equation \eqref{EX}) and $\tilde{\mathbf{X}}_t$ is the $t^\text{th}$ row of $\tilde{\mathbf{X}}$. Given these, the encoder DCGRU cell is formulated in Equation \eqref{enc_dcgru} as follows:
\begin{equation} \label{enc_dcgru}
\begin{split}
    \mathbf{r}^E_t &= \sigma_{\text{GRU}} (\boldsymbol{\zeta}^E_r \ast _{\mathcal{G}(\tilde{\mathbf{A}}_t,\tilde{\mathbf{D}}_t)} [\tilde{\mathbf{X}}_t, \mathbf{H}^E_{t-1}]\mathbf{W}^E_r + \mathbf{b}^E_r), \\
    \mathbf{u}^E_t &= \sigma_{\text{GRU}} (\boldsymbol{\zeta}^E_u \ast _{\mathcal{G}(\tilde{\mathbf{A}}_t,\tilde{\mathbf{D}}_t)} [\tilde{\mathbf{X}}_t, \mathbf{H}^E_{t-1}]\mathbf{W}^E_u + \mathbf{b}^E_u), \\
    \mathbf{C}^E_t &= \text{tanh} (\boldsymbol{\zeta}^E_C \ast _{\mathcal{G}(\tilde{\mathbf{A}}_t,\tilde{\mathbf{D}}_t)} [\tilde{\mathbf{X}}_t, (\mathbf{r}_t \odot \mathbf{H}^E_{t-1})]\mathbf{W}^E_C + \mathbf{b}^E_C), \\
    \mathbf{H}^E_t &= \mathbf{u}^E_t \odot \mathbf{H}^E_{t-1} + (1-\mathbf{u}^E_t) \odot \mathbf{C}^E_t,
\end{split}
\end{equation}
where $\mathbf{H}^E_{0} = 0$. The final hidden state of the encoders $\mathbf{H}^E_l$ is sent to the decoders as the initial decoder hidden state. For decoders, $t \in \{1,2, \ldots, h\}$. $\widehat{\mathbf{Y}}_{t}$ is the $t^\text{th}$ row of $\widehat{\mathbf{Y}}$. The decoder DCGRU cell is formulated in Equation \eqref{dec_dcgru} as follows:
\begin{equation} \label{dec_dcgru}
\begin{split}
    \mathbf{r}^D_t &= \sigma_{\text{GRU}} (\boldsymbol{\zeta}^D_r \ast _{\mathcal{G}(\tilde{\mathbf{A}}_{t+l},\tilde{\mathbf{D}}_{t+l})} [\widehat{\mathbf{Y}}_{t-1}, \mathbf{H}^D_{t-1}]\mathbf{W}^D_r + \mathbf{b}^D_r), \\
    \mathbf{u}^D_t &= \sigma_{\text{GRU}} (\boldsymbol{\zeta}^D_u \ast _{\mathcal{G}(\tilde{\mathbf{A}}_{t+l},\tilde{\mathbf{D}}_{t+l})} [\widehat{\mathbf{Y}}_{t-1}, \mathbf{H}^D_{t-1}]\mathbf{W}^D_u + \mathbf{b}^D_u), \\
    \mathbf{C}^D_t &= \text{tanh} (\boldsymbol{\zeta}^D_C \ast _{\mathcal{G}(\tilde{\mathbf{A}}_{t+l},\tilde{\mathbf{D}}_{t+l})} [\widehat{\mathbf{Y}}_{t-1}, (\mathbf{r}_t \odot \mathbf{H}^D_{t-1})]\mathbf{W}^D_C + \mathbf{b}^D_C), \\
    \mathbf{H}^D_t &= \mathbf{u}^D_t \odot \mathbf{H}^D_{t-1} + (1-\mathbf{u}^D_t) \odot \mathbf{C}^D_t,
\end{split}
\end{equation}
where $\widehat{\mathbf{Y}}_{0} = 0$ and $\mathbf{H}^D_{0} = \mathbf{H}^E_l$. For consistency, the same nonlinear activation function, $\sigma_{\text{GRU}}$, is used for both the update and reset gates in the encoders and decoders. For conciseness, the DCGRU cell of the encoders and decoders can be summarized in Equation \eqref{DCRNN-HAR enc cell} and Equation \eqref{DCRNN-HAR dec cell}, respectively.
\begin{equation} \label{DCRNN-HAR enc cell}
    \mathbf{H}^E_t = U_{\{\boldsymbol{\zeta}^E_r,\boldsymbol{\zeta}^E_u,\boldsymbol{\zeta}^E_C,\mathbf{W}^E_r,\mathbf{W}^E_u,\mathbf{W}^E_C,\mathbf{b}^E_r,\mathbf{b}^E_u,\mathbf{b}^E_C\}}(\tilde{\mathbf{X}}_t, \mathbf{H}^E_{t-1}, \tilde{\mathbf{A}}_t,\tilde{\mathbf{D}}_t),\forall t \in \{1,2,\ldots,l\}.
\end{equation}
\begin{equation} \label{DCRNN-HAR dec cell}
    \mathbf{H}^D_t = U_{\{\boldsymbol{\zeta}^D_r,\boldsymbol{\zeta}^D_u,\boldsymbol{\zeta}^D_C,\mathbf{W}^D_r,\mathbf{W}^D_u,\mathbf{W}^D_C,\mathbf{b}^D_r,\mathbf{b}^D_u,\mathbf{b}^D_C\}}(\widehat{\mathbf{Y}}_{t-1}, \mathbf{H}^D_{t-1}, \tilde{\mathbf{A}}_{t+l}, \tilde{\mathbf{D}}_{t+l}), \forall t \in \{1,2,\ldots,h\}.
\end{equation}
The DCGRU cell in the proposed DCRNN-HAR model processes the masked RV data sequentially, whereas the HAR component linearly aggregates the non-overlapping past daily, weekly and monthly RV patterns in parallel. Specifically, to generate the RV forecast $\widehat{\mathbf{Y}}$ based on masked inputs $\tilde{\mathbf{X}} \in \mathbb{R}^{l \times N}$, the past daily RV pattern is calculated as: $\tilde{\mathbf{X}}_d = \tilde{\mathbf{X}}_l$, the past weekly RV pattern is calculated as: $\tilde{\mathbf{X}}_w = \frac{1}{4}\sum_{j = l-4} ^ {l-1} \tilde{\mathbf{X}}_j$, and the past monthly RV pattern is calculated as: $\tilde{\mathbf{X}}_m = \frac{1}{17}\sum_{j = l-21} ^ {l-5} \tilde{\mathbf{X}}_j$. The outputs from the DCRNN and HAR components are summed together as shown in Equation \eqref{DCRNN-HAR forecast}. 
\begin{equation} \label{DCRNN-HAR forecast}
\begin{split}
    \widehat{\mathbf{X}}_{\text{HAR}} &= \alpha + \beta_d \tilde{\mathbf{X}}_d + \beta_w \tilde{\mathbf{X}}_w + \beta_m \tilde{\mathbf{X}}_m, \\
    \widehat{\mathbf{Y}}_{t} &= \mathbf{H}^D_t \mathbf{W}_{\text{out}} + \widehat{\mathbf{X}}_{\text{HAR}},
\end{split}
\end{equation}
where $\alpha$, $\beta_d$, $\beta_w$, $\beta_m$ and $\mathbf{W}_{\text{out}}$ are learnable parameters. All the learnable parameters of the DCRNN and HAR components of the proposed model are estimated simultaneously. The pseudo-code in Algorithm \ref{DCRNN-HAR pseudo-code} describes the detailed forecasting process of the DCRNN-HAR model.

\begin{algorithm}[H]
\caption{The forecasting process of the DCRNN-HAR model}\label{DCRNN-HAR pseudo-code}
\begin{algorithmic}[1]
    \STATE \textbf{Data:} input RV $\mathbf{X} \in \mathbb{R}^{l \times N}$, target RV $\mathbf{Y} \in \mathbb{R}^{h \times N}$, graph adjacency matrix $\mathbf{A} \in \mathbb{R}^{N \times N}$, $l=22$, $h=1,5,22$
    \STATE \textbf{Apply data masks:} $\tilde{\mathbf{X}} = \mathbf{E}^X \odot \mathbf{X}$, $ \tilde{\mathbf{Y}}=\mathbf{E}^Y \odot \mathbf{Y}$
    \STATE \textbf{Encoder initialization:} $t = 1$ ,$ \mathbf{H}^E_0 = 0$
    \STATE \textbf{Encoder:}\\
        \STATE \hspace{1.5em}\textbf{While} $t \leq l$ \textbf{do}
            \STATE \hspace{3em} $\tilde{\mathbf{A}}_t \leftarrow \mathbf{E}^{A_t} \odot \mathbf{A}$
            \STATE \hspace{3em} $\mathbf{H}^E_t \leftarrow U_{\{\boldsymbol{\zeta}^E_r,\boldsymbol{\zeta}^E_u,\boldsymbol{\zeta}^E_C,\mathbf{W}^E_r,\mathbf{W}^E_u,\mathbf{W}^E_C,\mathbf{b}^E_r,\mathbf{b}^E_u,\mathbf{b}^E_C\}}(\tilde{\mathbf{X}}_t, \mathbf{H}^E_{t-1}, \tilde{\mathbf{A}}_t,\tilde{\mathbf{D}}_t)$
            \STATE \hspace{3em} $t \leftarrow t+1$
        \STATE \hspace{1.5em}\textbf{End While}
        \STATE \textbf{HAR modeling:} $\widehat{\mathbf{X}}_{\text{HAR}} = \alpha + \beta_d \tilde{\mathbf{X}}_d + \beta_w \tilde{\mathbf{X}}_w + \beta_m \tilde{\mathbf{X}}_m$
        \STATE \textbf{Decoder initialization:} $t = 1$, $\widehat{\mathbf{Y}}_{0} = 0$, $\mathbf{H}^D_{0} = \mathbf{H}^E_l$, $\text{forecastList} = []$ 
        \STATE \textbf{Decoder:}
        \STATE \hspace{1.5em}\textbf{While} $t \leq h$ \textbf{do}
            \STATE \hspace{3em} $\tilde{\mathbf{A}}_{t+l} \leftarrow \mathbf{E}^{A_{t+l}} \odot \mathbf{A}$
            \STATE \hspace{3em} $\mathbf{H}^D_t \leftarrow U_{\{\boldsymbol{\zeta}^D_r,\boldsymbol{\zeta}^D_u,\boldsymbol{\zeta}^D_C,\mathbf{W}^D_r,\mathbf{W}^D_u,\mathbf{W}^D_C,\mathbf{b}^D_r,\mathbf{b}^D_u,\mathbf{b}^D_C\}}(\widehat{\mathbf{Y}}_{t-1}, \mathbf{H}^D_{t-1}, \tilde{\mathbf{A}}_{t+l}, \tilde{\mathbf{D}}_{t+l})$
            \STATE \hspace{3em} $\widehat{\mathbf{Y}}_{t} \leftarrow \mathbf{H}^D_t \mathbf{W}_{\text{out}} + \widehat{\mathbf{X}}_{\text{HAR}}$
            \STATE \hspace{3em} forecastList.append($\widehat{\mathbf{Y}}_{t}$)
            \STATE \hspace{3em} $t \leftarrow t+1$
        \STATE \hspace{1.5em}\textbf{End While}
        \STATE $\widehat{\mathbf{Y}}$ = \text{stack}(forecastList)
    \RETURN $\widehat{\mathbf{Y}}$
\end{algorithmic}
\end{algorithm}

\begin{figure}[H]
\centering
\includegraphics[width=\textwidth]{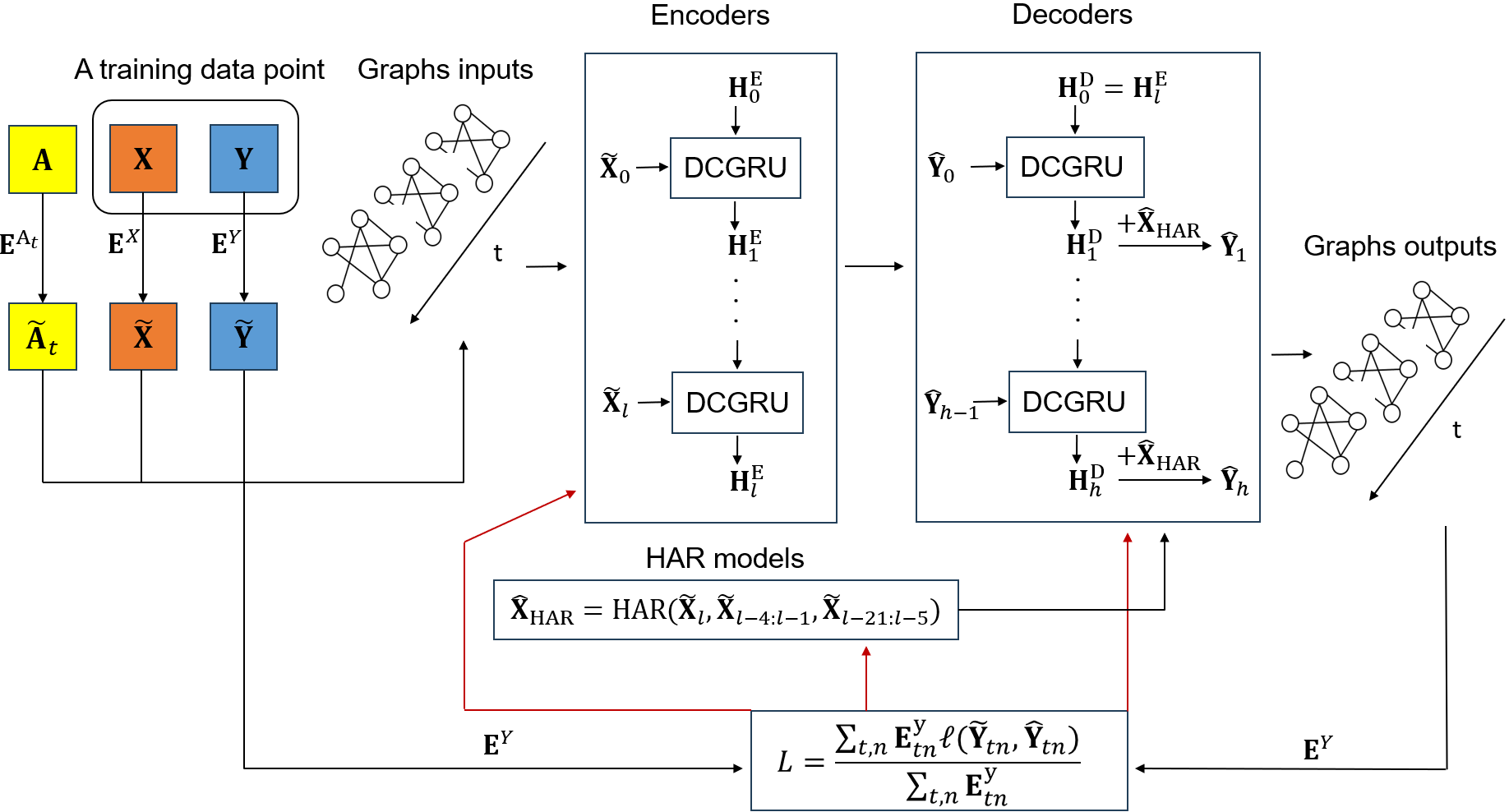}
\caption{An overview of the DCRNN-HAR model. }
\label{model_drawing}
\end{figure}

Regarding the loss measurement between the RV forecasts $\widehat{\mathbf{Y}}$ and the true future RV observations $\tilde{\mathbf{Y}}$, the forecasting loss is measured only on active trading days (Equation \eqref{L}). The parameters in the HAR components and the DCRNN components are updated all together to minimize the forecasting loss. An overview of the training process of the DCRNN-HAR model is shown in Figure \ref{model_drawing}. The black arrows indicate the forward propagation to generate RV forecasts and the red arrows represent the backward propagation to update the values of learnable parameters. Besides, a more detailed description of the data processing stage can be found in Figure \ref{data_prep_demo}.

\section{Empirical Study}\label{sec:empirical_study}
This section empirically evaluates the out-of-sample RV forecasting performance of the proposed DCRNN-HAR model with 8 representative global markets. Its performance is compared to the other 5 baseline models, including the HAR type and GNN-based models. Further discussions on the forecasting results are presented to offer a comprehensive analysis of the performance of the models and the RV dynamics of global stock markets. 

\subsection{Model Evaluation Tools}
To statistically assess the effectiveness of the DCRNN-HAR model, different evaluation criteria are employed. For comparison purposes, MSE and MAE loss functions are used to measure the RV forecasting accuracy of models on each stock market index. While the MAE provides a straightforward interpretation of average error magnitude and is less sensitive to extreme values, thereby offering robustness against outliers, the MSE is commonly used for its property of penalizing larger errors more severely. Besides, the market-wise Model Confidence Set (MCS) test \citep{Hansen2011} is conducted. Given a set of candidate forecasting models, MCS includes a subset of models that have statistically superior forecasting performances at a given confidence level (chosen as 75\%).

The hardware and software configurations are:
\begin{itemize}
    \item Operating system: Windows 11.
    \item CPU: 13th Gen Intel(R) Core(TM) i9-13900HX.
    \item GPU: NVIDIA GeForce RTX 4080 Laptop GPU.
    \item Software: Python 3.10.14; NumPy 1.26.4; PyTorch 1.13.1+cu116.
\end{itemize}

\subsection{Benchmark Datasets and Baseline Models}
The benchmark dataset covers the period from October 2006 to June 2022. It consists of $4079$ RV observations for each of the $8$ stock market indices on their union trading days. These indices are from both markets that are known to be influential and those that are not. The indices are SPX (US), GDAXI (Germany), FCHI (France), FTSE (UK), OMXSPI (Sweden), N225 (Japan), KS11 (South Korea) and HSI (Hong Kong), which are often used in RV forecasting studies (e.g., \citet{Liang2020}, \citet{Son2023}). The RV is calculated based on the $5$-minute high-frequency returns. Specifically, each value in the dataset is the square root of the corresponding RV data and is scaled by $100$. The dataset is partitioned into an in-sample dataset and an out-of-sample dataset. The in-sample dataset accounts for 70\% of the dataset from October 2006 to October 2017, while the out-of-sample dataset is from October 2017 to June 2022. Multiple forecasting windows are applied: $h = 1, 5, 22$, which represents the short-term (daily), mid-term (weekly) and long-term (monthly) forecasting. In addition, the look-back window $l$ is chosen to be $22$. All models are trained on the in-sample dataset. For each day of the out-of-sample period, the most recent $22$ RV observations (look-back window $l=22$) of each stock market index are loaded into the trained models to generate $h$-step-ahead forecasts. The forecast window then moves forward by one day and the procedure repeats for the entire out-of-sample period. The model training time is reported in Tables \ref{time MSE} and \ref{time MAE} and is discussed in Section \ref{Computational Time Discussion}.

\begin{figure}[H]
    \centering
    \includegraphics[width=16cm]{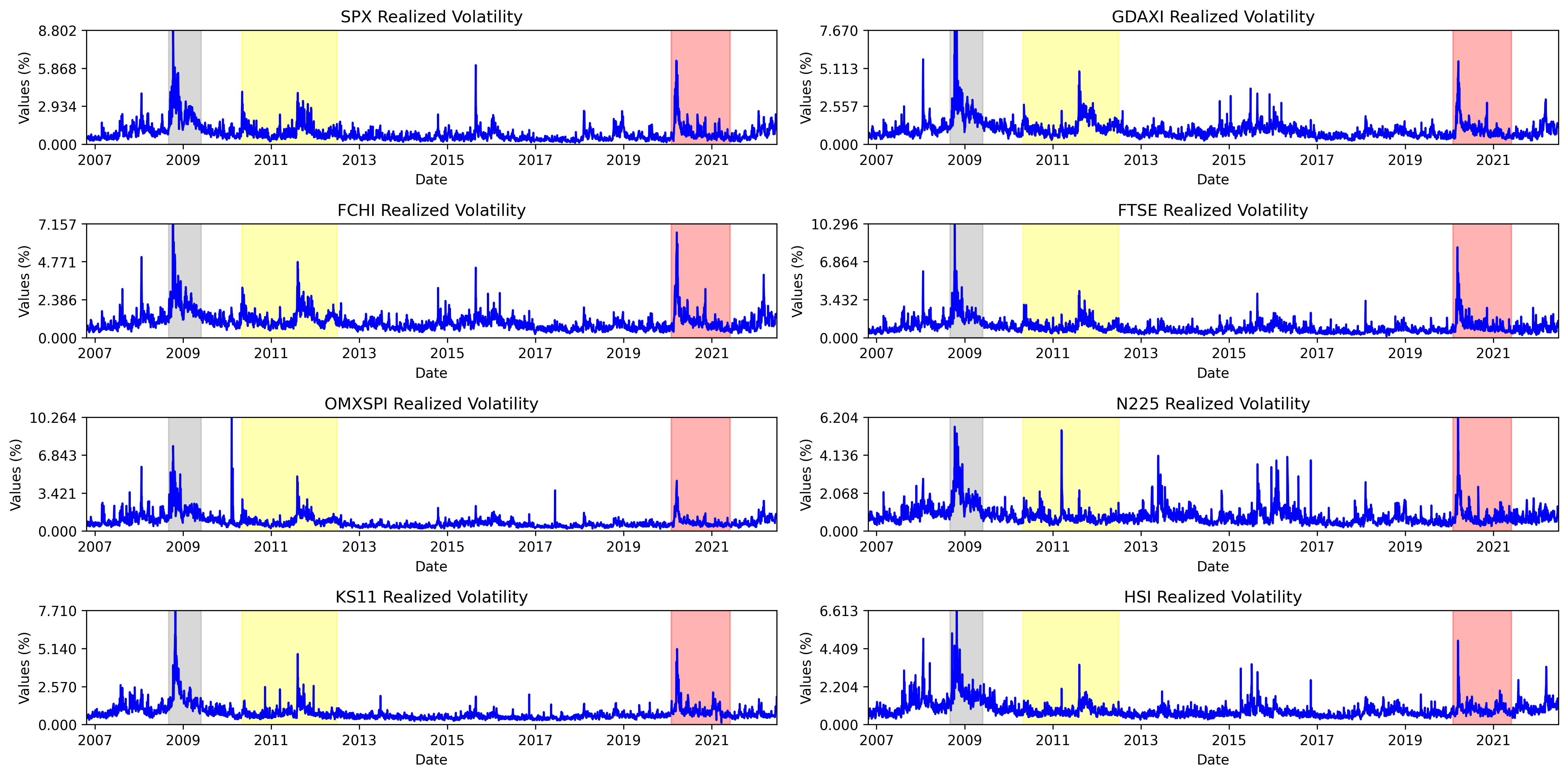}
    \caption{Visualization of the square-rooted RV series of each stock market index}
    \label{Visualize RV}
\end{figure}

The descriptive analysis of the square-rooted data for each market index is reported in the table below (Table \ref{tab:descriptive_stats}). The Augmented Dickey-Fuller (ADF) test \citep{MacKinnon1994, Cheung1995} is also conducted. At the $5\%$ significance level, the null hypothesis that a unit root exists in the square-rooted RV series of each stock market index is rejected for all time series, as the p-values are all less than $5\%$. This indicates that each square-rooted RV series is stationary. Besides, in Figure \ref{Visualize RV}, the scaled square-rooted RV series of each stock market index is visualized, where grey, yellow and red shadings highlight the Global Financial Crisis period, European Sovereign Debt Crisis period, and COVID-19 Pandemic period, respectively.
\begin{table}[H]
\centering
\begin{threeparttable}
\begin{tabularx}{0.98\textwidth}{lXXXXXX}
\toprule
Index & T & Mean & Standard Deviation & Skewness & Kurtosis  & {ADF \text{p-value}} \\
\midrule
   SPX & 4079 & 0.808 &    0.676 &     3.257 &    17.890 &    0.000 \\
 GDAXI & 4079 & 0.940 &    0.608 &     3.005 &    17.515 &    0.000 \\
  FCHI & 4079 & 0.956 &    0.595 &     2.973 &    17.072 &    0.000 \\
  FTSE & 4079 & 0.903 &    0.641 &     3.737 &    28.513 &    0.000 \\
OMXSPI & 4079 & 0.792 &    0.608 &     4.228 &    36.349 &    0.000 \\
  N225 & 4079 & 0.768 &    0.534 &     2.935 &    16.423 &    0.000 \\
  KS11 & 4079 & 0.729 &    0.520 &     3.704 &    26.060 &    0.000 \\
   HSI & 4079 & 0.810 &    0.509 &     2.838 &    17.278 &    0.000 \\
\bottomrule
\end{tabularx}
\end{threeparttable}
\caption{Descriptive statistics of the scaled square-rooted RV data.}
\label{tab:descriptive_stats}
\end{table}

In addition to the proposed model, this study considers the following baseline models: HAR, VHAR, HAR-KS, GNN-HAR, and STG-Spillover, as reviewed in Section \ref{sec:literature_review}. Each model is independently trained in the in-sample dataset and then used to generate $h$-step-ahead forecasts over the out-of-sample period. This setup allows a direct comparison of forecasting performances across different horizons $h$ and stock markets.

The RV dataset used in the experiment, the code of the experiment and the relevant hyperparameter settings can be found at {\small\url{https://github.com/MikeZChi/DCRNN-HAR.git}}.

\subsection{Forecasting Results}\label{Forecasting Results}
The MSE and MAE scores of the forecast series generated by each model on out-of-sample data under different forecast horizons are listed in Tables \ref{MSE h=1} to \ref{MAE h=22}. For each market, the model with the smallest forecasting error score is highlighted in blue. In addition, the MCS test on the $75\%$ confidence level is also conducted for each forecasting setting. The set of superior models that are included in MCS is highlighted in gray shading, as shown in Tables \ref{MSE h=1} to \ref{MAE h=22}.
\begin{table}[H]
    \centering
    \begin{tabular}{lcccccc}
        \toprule
        {$h=1$} & HAR       & VHAR      & HAR-KS    & GNN-HAR    & STG-Spillover & DCRNN-HAR     \\
        \midrule
        SPX     & 0.144      & 0.144      & 0.143      & \cg{0.136} & 0.140          & \cb{\cg{0.125}} \\
        GDAXI   & \cg{0.094} & \cg{0.088} & \cg{0.089} & \cg{0.090} & 0.096          & \cb{\cg{0.083}} \\
        FCHI    & 0.124      & \cg{0.117} & \cg{0.118} & \cg{0.118} & 0.120          & \cb{\cg{0.104}} \\
        FTSE    & 0.183      & 0.175      & 0.174      & \cg{0.170} & 0.171          & \cb{\cg{0.163}} \\
        OMXSPI  & 0.061      & 0.063      & 0.065      & \cg{0.055} & \cg{0.057}     & \cb{\cg{0.052}} \\
        N225    & 0.098      & 0.089      & 0.088      & 0.089      & 0.086          & \cb{\cg{0.074}} \\
        KS11    & 0.072      & 0.067      & 0.068      & 0.064      & 0.066          & \cb{\cg{0.053}} \\
        HSI     & 0.073      & 0.075      & 0.075      & \cg{0.065} & \cg{0.066}     & \cb{\cg{0.063}} \\
        \bottomrule
    \end{tabular}
    \caption{MSE comparison at short-term forecasting horizon ($h = 1$). Cells highlighted in gray represent models selected by the MCS test; numbers highlighted in blue indicate the best-performing model per index.}
    \label{MSE h=1}
\end{table}

\begin{table}[H]
    \centering
    \begin{tabular}{lcccccc}
        \toprule
        {$h=1$} & HAR       & VHAR      & HAR-KS    & GNN-HAR    & STG-Spillover & DCRNN-HAR     \\
        \midrule
        SPX     & 0.227      & 0.224      & 0.224      & 0.220      & 0.218          & \cb{\cg{0.207}} \\
        GDAXI   & 0.195      & 0.192      & 0.194      & 0.188      & 0.190          & \cb{\cg{0.178}} \\
        FCHI    & 0.221      & 0.219      & 0.219      & 0.212      & 0.213          & \cb{\cg{0.201}} \\
        FTSE    & 0.240      & 0.239      & 0.237      & 0.232      & 0.231          & \cb{\cg{0.221}} \\
        OMXSPI  & 0.154      & 0.158      & 0.159      & 0.152      & 0.147          & \cb{\cg{0.141}} \\
        N225    & 0.182      & 0.178      & 0.176      & 0.178      & 0.171          & \cb{\cg{0.160}} \\
        KS11    & 0.156      & 0.150      & 0.151      & 0.147      & 0.148          & \cb{\cg{0.133}} \\
        HSI     & 0.163      & 0.165      & 0.163      & 0.153      & 0.159          & \cb{\cg{0.147}} \\
        \bottomrule
    \end{tabular}
    \caption{MAE comparison at short-term forecasting horizon ($h = 1$). Cells highlighted in gray represent models selected by the MCS test; numbers highlighted in blue indicate the best-performing model per index.}
    \label{MAE h=1}
\end{table}

\begin{table}[H]
    \centering
    \begin{tabular}{lcccccc}
        \toprule
        {$h=5$} & HAR       & VHAR      & HAR-KS    & GNN-HAR    & STG-Spillover & DCRNN-HAR     \\
        \midrule
        SPX     & \cg{0.248} & \cg{0.249} & \cg{0.251} & \cg{0.216} & \cg{0.192}     & \cb{\cg{0.158}} \\
        GDAXI   & 0.153      & 0.162      & 0.149      & 0.136      & \cg{0.134}     & \cb{\cg{0.095}} \\
        FCHI    & 0.201      & 0.206      & 0.196      & 0.184      & 0.165          & \cb{\cg{0.122}} \\
        FTSE    & \cg{0.229} & \cg{0.239} & \cg{0.226} & \cg{0.216} & \cg{0.211}     & \cb{\cg{0.169}} \\
        OMXSPI  & 0.147      & 0.158      & 0.148      & 0.136      & 0.084          & \cb{\cg{0.059}} \\
        N225    & \cg{0.125} & \cg{0.126} & \cg{0.117} & \cg{0.119} & \cg{0.103}     & \cb{\cg{0.087}} \\
        KS11    & \cg{0.104} & \cg{0.101} & \cg{0.097} & \cg{0.092} & \cg{0.071}     & \cb{\cg{0.067}} \\
        HSI     & \cg{0.097} & \cg{0.098} & \cg{0.094} & \cg{0.090} & \cg{0.075}     & \cb{\cg{0.065}} \\
        \bottomrule
    \end{tabular}
    \caption{MSE comparison at mid-term forecasting horizon ($h = 5$). Cells highlighted in gray represent models selected by the MCS test; numbers highlighted in blue indicate the best-performing model per index.}
    \label{MSE h=5}
\end{table}

\begin{table}[H]
    \centering
    \begin{tabular}{lcccccc}
        \toprule
        {$h=5$} & HAR       & VHAR      & HAR-KS    & GNN-HAR    & STG-Spillover & DCRNN-HAR     \\
        \midrule
        SPX     & 0.284      & 0.283      & 0.281      & 0.273      & 0.245          & \cb{\cg{0.226}} \\
        GDAXI   & 0.236      & 0.241      & 0.235      & 0.227      & 0.210          & \cb{\cg{0.187}} \\
        FCHI    & 0.265      & 0.270      & 0.264      & 0.254      & 0.236          & \cb{\cg{0.210}} \\
        FTSE    & 0.268      & 0.271      & 0.266      & 0.261      & 0.253          & \cb{\cg{0.224}} \\
        OMXSPI  & 0.193      & 0.202      & 0.194      & 0.185      & 0.166          & \cb{\cg{0.145}} \\
        N225    & 0.202      & 0.213      & 0.197      & 0.196      & 0.188          & \cb{\cg{0.169}} \\
        KS11    & 0.179      & 0.177      & 0.176      & 0.171      & 0.165          & \cb{\cg{0.145}} \\
        HSI     & 0.185      & 0.185      & 0.183      & 0.185      & 0.160          & \cb{\cg{0.151}} \\
        \bottomrule
    \end{tabular}
    \caption{MAE comparison at mid-term forecasting horizon ($h = 5$). Cells highlighted in gray represent models selected by the MCS test; numbers highlighted in blue indicate the best-performing model per index.}
    \label{MAE h=5}
\end{table}

\begin{table}[H]
    \centering
    \begin{tabular}{lcccccc}
        \toprule
        {$h=22$}& HAR       & VHAR      & HAR-KS    & GNN-HAR    & STG-Spillover & DCRNN-HAR     \\
        \midrule
        SPX     & \cg{0.428} & \cg{0.389} & 0.425      & \cg{0.410} & \cg{0.204}     & \cb{\cg{0.195}} \\
        GDAXI   & \cg{0.256} & \cg{0.268} & \cg{0.263} & \cg{0.254} & \cg{0.129}     & \cb{\cg{0.117}} \\
        FCHI    & \cg{0.332} & \cg{0.333} & \cg{0.336} & \cg{0.331} & \cg{0.163}     & \cb{\cg{0.146}} \\
        FTSE    & \cg{0.365} & \cg{0.351} & \cg{0.363} & \cg{0.365} & \cg{0.207}     & \cb{\cg{0.191}} \\
        OMXSPI  & 0.202      & 0.214      & 0.208      & 0.200      & 0.090          & \cb{\cg{0.070}} \\
        N225    & \cg{0.179} & 0.195      & \cg{0.188} & \cg{0.184} & \cg{0.107}     & \cb{\cg{0.101}} \\
        KS11    & \cg{0.146} & 0.154      & 0.158      & \cg{0.144} & \cg{0.083}     & \cb{\cg{0.079}} \\
        HSI     & 0.116      & 0.118      & 0.122      & 0.118      & 0.063          & \cb{\cg{0.059}} \\
        \bottomrule
    \end{tabular}
    \caption{MSE comparison at long-term forecasting horizon ($h = 22$). Cells highlighted in gray represent models selected by the MCS test; numbers highlighted in blue indicate the best-performing model per index.}
    \label{MSE h=22}
\end{table}

\begin{table}[H]
    \centering
    \begin{tabular}{lcccccc}
        \toprule
        {$h=22$}& HAR       & VHAR      & HAR-KS    & GNN-HAR    & STG-Spillover & DCRNN-HAR     \\
        \midrule
        SPX     & 0.357      & 0.334      & 0.351      & 0.359      & 0.258          & \cb{\cg{0.251}} \\
        GDAXI   & 0.275      & 0.286      & 0.277      & 0.270      & 0.209          & \cb{\cg{0.200}} \\
        FCHI    & 0.314      & 0.324      & 0.317      & 0.310      & 0.228          & \cb{\cg{0.223}} \\
        FTSE    & 0.311      & 0.318      & 0.315      & 0.314      & 0.238          & \cb{\cg{0.234}} \\
        OMXSPI  & 0.222      & 0.229      & 0.225      & 0.226      & 0.166          & \cb{\cg{0.156}} \\
        N225    & 0.231      & 0.244      & 0.237      & 0.240      & 0.184          & \cb{\cg{0.180}} \\
        KS11    & 0.208      & 0.207      & 0.208      & 0.209      & 0.169          & \cb{\cg{0.164}} \\
        HSI     & 0.205      & 0.205      & 0.209      & 0.200      & 0.154          & \cb{\cg{0.151}} \\
        \bottomrule
    \end{tabular}
    \caption{MAE comparison at long-term forecasting horizon ($h = 22$). Cells highlighted in gray represent models selected by the MCS test; numbers highlighted in blue indicate the best-performing model per index.}
    \label{MAE h=22}
\end{table}

In addition, a MCS test summary is presented in Table \ref{MCS summary}. For each model and each market, the number of times that a model is included in MCS for the 3 different forecasting horizons and 2 loss measurements is presented. The ``Total'' row shows the sum of the MCS count for each model. According to the out-of-sample forecast evaluation results from Tables \ref{MSE h=1} to \ref{MAE h=22} and the MCS summary table, the proposed DCRNN-HAR model consistently generates the most accurate RV forecasting series and always remains in the MCS at the $75\%$ confidence level, regardless of the choice of target stock markets, forecast horizons or evaluation criteria. This demonstrates the robustness of the proposed DCRNN-HAR model. In contrast, the baseline models in comparison generate less accurate RV forecasts and stay in the MCS much less frequently.

\begin{table}[H]
    \centering
    \begin{tabular}{lcccccc}
        \toprule
               & HAR & VHAR & HAR-KS & GNN-HAR & STG-Spillover & DCRNN-HAR \\
        \midrule
        SPX     & 2   & 2    & 1      & 3      & 2             & 6        \\
        GDAXI   & 2   & 2    & 2      & 2      & 2             & 6        \\
        FCHI    & 1   & 2    & 2      & 2      & 1             & 6        \\
        FTSE    & 2   & 2    & 2      & 3      & 2             & 6        \\
        OMXSPI  & 0   & 0    & 0      & 1      & 1             & 6        \\
        N225    & 2   & 1    & 2      & 2      & 2             & 6        \\
        KS11    & 2   & 1    & 1      & 2      & 2             & 6        \\
        HSI     & 1   & 1    & 1      & 2      & 2             & 6        \\
        \midrule
        \textbf{Total} & \textbf{12} & \textbf{11} & \textbf{11} & \textbf{17} & \textbf{14} & \textbf{48} \\
        \bottomrule
    \end{tabular}
    \caption{The summary table of the MCS test. For each model and each market, the number of times that a model is included in MCS for the 3 different forecasting horizons and 2 loss measurements is presented. The ``Total'' row shows the sum of MCS count for each model.}
    \label{MCS summary}
\end{table}

\subsection{Net Volatility Spillover and RV Forecasting}
In this section, we aim to link the net volatility spillover and RV forecasting results of different models and markets, to provide further insights on the performance of the models linked to the behaviors of different types of stock markets (influential and impressionable). 

In Section \ref{sec:volatility_interconnection}, the volatility interconnection graph is constructed under the DY framework and its adjacency matrix is calculated as $\mathbf{A} = [\tilde{\boldsymbol{\theta}}^g(h)]^\top$. According to \citet{Diebold2012}, the net volatility spillover from stock market $i$ to all other stock markets is the difference between: 1) the volatility spillovers transmitted by stock market $i$ \emph{to} other stock markets $j$, and 2) the volatility spillovers received by stock market $i$ \emph{from} other stock markets $j$. The volatility spillovers transmitted by stock market $i$ to others can be denoted as $S^g_{\cdot i}$ and is calculated in Equation \eqref{vsp_transmit_i}. On the other hand, the volatility spillovers received by stock market $i$ from others can be denoted as $S^g_{i\cdot}$ and is calculated in Equation \eqref{vsp_receive_i}.
\begin{equation} \label{vsp_transmit_i}
    S^g_{\cdot i}(h) = 100\frac{\sum_{j=1, j \neq i}^N \tilde{\boldsymbol{\theta}}_{ji}^g(h)}{N}.
\end{equation}
\begin{equation} \label{vsp_receive_i}
    S^g_{i\cdot}(h) = 100\frac{\sum_{j=1, j \neq i}^N \tilde{\boldsymbol{\theta}}_{ij}^g(h)}{N}.
\end{equation}
Here, $N$ is the number of stock markets included in this research: $N = 8$. Thus, the net volatility spillover from stock market $i$ can be calculated in Equation \eqref{net_vsp_i}. The choice of $h$ aligns with the forecast window settings: $h = 1, 5, 22$. The net volatility spillover calculation results are included in Tables \ref{net_vsp_table_1}, \ref{net_vsp_table_5} and \ref{net_vsp_table_22}.
\begin{equation} \label{net_vsp_i}
    S^g_{i}(h) = S^g_{\cdot i}(h) - S^g_{i\cdot}(h).
\end{equation}

\begin{table}[H]
    \centering
    \begin{tabular}{lcccccccc}
        \toprule
        {} & SPX & GDAXI & FCHI & FTSE & OMXSPI & N225 & KS11 & HSI \\
        \midrule
        {$S^g_{i}(1)$} & 2.408 & 2.221 & 2.848 & 1.326 & -2.695 & -3.101 & -1.472 & -2.536 \\
        \bottomrule
    \end{tabular}
    \caption{Net volatility spillover from each stock market to all other stock markets at short-term forecasting horizon ($h=1$).}
    \label{net_vsp_table_1}
\end{table}

\begin{table}[H]
    \centering
    \begin{tabular}{lcccccccc}
        \toprule
        {} & SPX & GDAXI & FCHI & FTSE & OMXSPI & N225 & KS11 & HSI \\
        \midrule
        {$S^g_{i}(5)$} & 5.105 & 1.688 & 2.763 & 1.526 & -1.943 & -4.425 & -1.633 & -4.081 \\
        \bottomrule
    \end{tabular}
    \caption{Net volatility spillover from each stock market to all other stock markets at mid-term forecasting horizon ($h=5$).}
    \label{net_vsp_table_5}
\end{table}

\begin{table}[H]
    \centering
    \begin{tabular}{lcccccccc}
        \toprule
        {} & SPX & GDAXI & FCHI & FTSE & OMXSPI & N225 & KS11 & HSI \\
        \midrule
        {$S^g_{i}(22)$} & 9.596 & 0.527 & 3.226 & 2.302 & -1.462 & -6.501 & -4.513 & -6.175 \\
        \bottomrule
    \end{tabular}
    \caption{Net volatility spillover from each stock market to all other stock markets at long-term forecasting horizon ($h=22$).}
    \label{net_vsp_table_22}
\end{table}
As shown in Tables \ref{net_vsp_table_1}, \ref{net_vsp_table_5} and \ref{net_vsp_table_22}, the signs of the values are fixed, despite the variation in their scale. Influential stock markets are those with positive net volatility spillover, which means they transmit more volatility shocks to other markets than they receive. In contrast, impressionable markets have negative net spillover because they receive greater volatility influence from external markets than they pass on. According to the MCS test summary Table \ref{MCS summary}, the baseline models are more likely to be included in the MCS when forecasting RV for influential markets. This implies that the performance of the baseline models is not consistent across different markets: they can generate accurate predictions for influential stock markets but struggle with forecasting RV for impressionable markets. This inconsistency shows that the baseline models do not have sufficient power to capture the volatility spillover effect. For indices in influential stock markets, their volatility pattern is dominated by internal dynamics and is less influenced by external factors, which is easier for the baseline models to learn. On the contrary, impressionable markets tend to absorb external shocks due to their exposure to inter-market influences, thus complicating their volatility dynamics. This highlights the asymmetric nature of the volatility spillover effect which is challenging for the baseline models. Meanwhile, with different masks and the combination of the DCRNN component and the HAR framework, the DCRNN-HAR model can process both cross-sectional volatility interrelationships and volatility auto-regressive dependencies flexibly. Thus, it is capable of handling the volatility dynamic of both influential and impressionable markets and has achieved the best forecasting performance as shown in Section \ref{Forecasting Results}. 

\subsection{Computational Time Discussion} \label{Computational Time Discussion}
This section discusses the computational cost of the proposed DCRNN-HAR and its comparison to other baseline models, to lend evidence on its practical applicability. Tables \ref{time MSE} and \ref{time MAE} report the computational time (in seconds) for each model trained under different objective functions (MSE and MAE) and across various forecasting horizons ($h=1,5,22$). As can be seen, the HAR-type models without the neural network component can be trained within seconds. For the GNN-HAR, STG-Spillover and DCRNN-HAR models, the inclusion of neural network components increases the training times to minutes, which is still reasonable given that the forecasting task is on a daily basis. Compared to the HAR-type models, the extra computational time of DCRNN-HAR leads to significantly improved RV forecasting accuracy. Meanwhile, the computational time of DCRNN-HAR is between that of the GNN-HAR and STG-Spillover models, which means that the enhanced forecasting power of DCRNN-HAR does not result in significantly increased additional training costs compared to other GNN-based models. Given the superior forecasting performance of DCRNN-HAR and the reasonable training cost, the proposed model serves as an attractive option for volatility forecasting of global stock markets.

\begin{table}[H]
    \centering
    \begin{tabular}{lcccccc}
        \toprule
        {MSE} & HAR & VHAR & HAR-KS & GNN-HAR & STG-Spillover & DCRNN-HAR \\
        \midrule
        $h=1$  & 0.064 & 0.117 & 0.073 & 104.993 & 383.062   & 269.474 \\
        $h=5$  & 0.091 & 0.132 & 0.093 & 94.114  & 353.026   & 286.301 \\
        $h=22$ & 0.072 & 0.147 & 0.099 & 87.284  & 534.250   & 441.251 \\
        \bottomrule
    \end{tabular}
    \caption{Computational time (in seconds) for different models under the MSE objective function.}
    \label{time MSE}
\end{table}

\begin{table}[H]
    \centering
    \begin{tabular}{lcccccc}
        \toprule
        {MAE} & HAR & VHAR & HAR-KS & GNN-HAR & STG-Spillover & DCRNN-HAR \\
        \midrule
        $h=1$  & 0.136 & 3.794 & 1.551 & 146.313 & 396.658   & 294.164 \\
        $h=5$  & 0.197 & 4.209 & 1.675 & 116.252 & 422.867   & 289.613 \\
        $h=22$ & 0.149 & 5.301 & 1.603 & 104.247 & 508.270   & 390.267 \\
        \bottomrule
    \end{tabular}
    \caption{Computational time (in seconds) for different models under the MAE objective function.}
    \label{time MAE}
\end{table}

\section{Conclusion}\label{sec:conclusion}
This paper aims to forecast the global stock market RV by incorporating dynamic graphs and both the common and uncommon trading days. The proposed DCRNN-HAR model uses novel data and graph masks to accommodate the different trading schedules of various stock markets. These mask designs allow the proposed model to generate forecasts on all trading days for different stock markets. Thus, the DCRNN-HAR model has a high practical utility for investors who invest globally or consider the influence of all other global markets while investing in a given local market. In addition, these designs help capture the volatility spillover effect and the changing dynamics of the volatility interconnection network. In the empirical study section, the DCRNN-HAR model generates the most accurate forecasts for all stock market indices under all forecasting settings.

This study not only contributes to RV forecasting for global stock market indices, but also suggests promising avenues for future research. While the literature offers several methods for constructing volatility interconnection graphs \citep{Son2023, Zhang2025}, there is a notable lack of in-depth analysis comparing the benefits and limitations of these various approaches. Addressing this gap could provide clearer insights into the effectiveness of different settings for modeling volatility interactions and ultimately lead to the development of more robust volatility analysis frameworks.

Hence, it is feasible to conclude that the proposed DCRNN-HAR model successfully integrates all trading days and the dynamic relationship network and generates more accurate RV forecasts. The DCRNN-HAR model is beneficial for understanding and capturing the volatility dynamic for different stock markets.

\section{Disclosure of Interest}
No conflict of interest is to be declared.

\section{Disclosure of Funding}
No funding was received.

\section{Data Availability Statement}
Data were downloaded from Oxford-man Institute's realized library. The authors confirm that the data supporting the findings of this study are available within the supplementary materials of the paper.

\newpage
\bibliographystyle{chicago}

\end{document}